\documentclass[floatfix,aps,pre,twocolumn,superscriptaddress,noeprint]{revtex4-1}

\usepackage{graphicx}
\usepackage{amsmath}
\usepackage{amssymb}
\usepackage{booktabs}
\usepackage{xcolor}
\usepackage[utf8]{inputenc}
\usepackage{natbib}
\usepackage{bm}
\usepackage{tabularx}   
\usepackage{MnSymbol,wasysym}
\usepackage{comment}

\begin{document}

\newcommand{\Jeff}[1]{{\textcolor{red}{#1}}}
\newcommand{\Derek}[1]{{\textcolor{green}{#1}}}
\newcommand{\Rev}[1]{{\textcolor{blue}{#1}}}

\title{Stress and stretching regulate dispersion in viscoelastic porous media flows}

\author{Manish Kumar}
\thanks{These authors contributed equally.}
\affiliation{Department of Mechanical Engineering, Purdue University, 585 Purdue Mall, West Lafayette, Indiana 47907, USA}
\author{Derek~M. Walkama}
\thanks{These authors contributed equally.}
\affiliation{Department of Mechanical Engineering, Tufts University, 200 College Avenue, Medford, Massachusetts 02155, USA}
\affiliation{Department of Physics and Astronomy, Tufts University, 574 Boston Avenue, Medford, Massachusetts 02155, USA}
\author{Arezoo~M. Ardekani}
\affiliation{Department of Mechanical Engineering, Purdue University, 585 Purdue Mall, West Lafayette, Indiana 47907, USA}
\author{Jeffrey~S. Guasto}
\thanks{Corresponding author: Jeffrey.Guasto@tufts.edu}
\affiliation{Department of Mechanical Engineering, Tufts University, 200 College Avenue, Medford, Massachusetts 02155, USA}

\begin{abstract}
In this work, we study the role of viscoelastic instability in the mechanical dispersion of fluid flow through porous media at high P\'eclet numbers. 
Using microfluidic experiments and numerical simulations, we show that viscoelastic instability in flow through a hexagonally ordered (staggered) medium strongly enhances dispersion transverse to the mean flow direction with increasing Weissenberg number (Wi).
In contrast, preferential flow paths can quench the elastic instability in disordered media, which has two important consequences for transport: 
First, the lack of chaotic velocity fluctuations reduces transverse dispersion relative to unstable flows. 
Second, the amplification of flow along preferential paths with increasing Wi causes strongly-correlated stream-wise flow that enhances longitudinal dispersion. 
Finally, we illustrate how the observed dispersion phenomena can be understood through the lens of Lagrangian stretching manifolds, which act as advective transport barriers and coincide with high stress regions in these viscoelastic porous media flows.
\end{abstract}


\maketitle
\section{Introduction}
The flow of viscoelastic fluids through porous media governs material transport and mixing in a range of geophysical, biological, and industrial systems~\cite{Kumar2022review,Datta2022}.
Bacterial biofilms proliferate in soils and cause infections in bodily tissue~\cite{Hall-Stoodley2004}, and filtration media are used in food and polymer processing~\cite{Anguiano2017}.
Polymer additives improve the efficacy of hydraulic fracturing and enhanced oil recovery (EOR)~\cite{Delshad2008, Wang2011, Wever2011}, including the remediation of oil ganglia~\cite{Rodriguez1993, De2018, Zamani2015, Aliabadian2020,Aramideh2019surfactant}. 
In the latter case of EOR for example, despite extensive efforts to observe and understand the impact of viscoelastic flow, no globally accepted remediation mechanism via polymer additives has been established~\cite{Fan2018, Wei2014, Haward2003,Zaitoun1998}.  
However, the onset of unsteady velocity fluctuations in such viscoelastic porous media flows~\cite{Scholz2014, Clarke2015, Clarke2016a} appears to play a critical role in microscale transport~\cite{Scholz2014, Jacob2017, Babayekhorasani2016, Aramideh2019}, where porous microstructure couples pore-scale viscoelastic flows~\cite{Browne2019} to sample-scale transport properties~\cite{bear1988}.
The non-Newtonian rheology of viscoelastic fluids encodes a memory of the flow history, whose non-trivial dependence on pore geometry~\cite{Kumar2021tristability,Kumar2022asymmetric,Kumar2022hysteresis,Kumar2021multistability} can result in viscoelastic instability~\cite{Kenney2013, Grilli2013, Scholz2014, Howe2015, Shi2016, Varshney2017, Qin2017, Walkama2020, Browne2021, Haward2021}.
A deeper understanding of the interplay between rheology, flow structure, and dispersion is paramount to predicting material transport in viscoelastic porous media flows.\par

In the absence of inertia, strong elastic stresses cause viscoelastic flow instabilities in porous media, which are heavily dependent on the flow geometry~\cite{Walkama2020, Haward2021}.
The transition to chaotic dynamics in viscoelastic flows is characterized by the Weissenberg number, $\textrm{Wi} = \tau\dot{\gamma}$, which compares elastic forces to viscous forces.  
Here, $\tau$ is the fluid relaxation time and $\dot{\gamma}=U/d$ is the characteristic shear rate, where $U$ is the average flow speed and $d$ is the characteristic obstacle diameter. 
Chaotic velocity fluctuations at large Wi have been shown to enhance transverse dispersion in ordered porous media flows~\cite{Scholz2014} via a ``lane-changing'' effect~\cite{De2017}. 
In contrast, markedly weaker dispersion enhancement has been reported for viscoelastic flows in disordered media~\cite{Aramideh2019, Babayekhorasani2016, Jacob2017}.
Recent experiments have shed new light on the geometry-dependent transition to chaos and the resulting flow topologies, which ultimately regulate the dispersion properties.
The critical Weissenberg number, $\textrm{Wi}_{cr}$, is highly sensitive to both the disorder of the medium~\cite{Walkama2020} and orientation of ordered media relative to the flow~\cite{Haward2021}.
Preferential flow paths in disordered media and along lattice directions in periodic media reduce extensional deformation and stress, and ultimately suppress the transition to chaos compared to staggered obstacle arrangements in ordered systems at the same $\textrm{Wi}$~\cite{Walkama2020, Haward2021}.
This topological and dynamical shift in the flow field with geometry must be intrinsically linked to the transport properties.
However, a comprehensive understanding of how viscoelastic flow instabilities regulate dispersion in porous media remains lacking.\par

In this work, we characterize how changes in viscoelastic flow topology regulate anisotropic dispersion in porous media flows, and elucidate the underlying mechanisms for geometry-dependent transport. 
Microfluidic experiments are complemented by numerical simulations for quasi-two-dimensional model porous media for both hexagonally ordered (staggered) and disordered arrays of cylindrical pillars.
Our results capture the accepted enhancement of transverse dispersion via flow instability in ordered flow and also reveal that the amplification of preferential flow paths in the disordered media increases longitudinal dispersion~[Fig.~\ref{fig:VEDisp_fig1}]. 
Furthermore, recent theoretical work based on a Lagrangian analysis of viscoelastic flows demonstrated that the fluid stretching field closely reflects the polymeric stress topology~\cite{Kumar2023PNAS}.
We show that transverse and longitudinal dispersion can be understood through the structure of the Lagrangian stretching field, whose manifolds act as barriers to advection and dynamically guide transport in both steady and unsteady flows.
These results demonstrate a potential mechanism for tuning anisotropic dispersion, and they illustrate a direct link between elastic stress and transport in viscoelastic porous media flows.

\section{Methods}
\subsection{Experimental methods}
Following established approaches from previous work~\cite{Walkama2020}, microfluidic devices were designed and fabricated with 25~mm long, 4~mm wide, and 50~$\mu$m high straight main channels, which contain arrays of cylindrical pillars (diameter, $d=50$~$\mu$m) in both an ordered and disordered configuration.
Photolithography masks were generated by first specifying an ordered, hexagonal array in a staggered orientation relative to the flow direction~\cite{Haward2021}, which had a lattice constant, $a=120$~$\mu$m.
The disordered geometry was created by randomly perturbing the pillar locations from the original lattice within a hexagonal circumradius, $a$.
The viscoelastic fluid is a solution of high molecular weight polyacrylamide (PAA; $18\times10^6$~MW) at a concentration of 150~ppm of PAA in a viscous Newtonian solvent (97\% aqueous glycerol)~\cite{Faustino2015}.
The solution was prepared by mixing 1~g of PAA into 200~mL of DI water using a magnetic stirrer for 1~hr. 
3~g of the aqueous PAA solution was mixed with 97~g of glycerol for 12~hr. 
Finally, the fluid was seeded with 0.5~$\mu$m and 1~$\mu$m diameter tracer particles for simultaneous particle image velocimetry (PIV) and particle tracking, respectively.\par

Capillary breakup extensional rheology (CaBER) was used to characterize the (longest) relaxation time, $\tau$, of the PAA solutions~\cite{Anna2001}, which provides a more relevant measure of the relaxation time for stongly elongational flows compared to shear rheology~\cite{Rothstein1999,Rothstein2001}. 
The PAA solution was stretched between two dowels, and the measured exponential decay rate of the liquid bridge diameter ($3\tau$) gave a relaxation time $\tau=1.14\pm0.1$ ($N=6$).
The shear-rate dependence of the viscoelastic fluid was characterized using a strain controlled rheometer (TA-2000) with a cone and plate geometry.
The polymer solution was pre-sheared at a rate of 1~s$^{-1}$ for 120~s, then each measurement was held at the respective shear rate for 60~s and measured for 15~s.
The PAA solution exhibited a weak shear thinning behavior, which is well fitted by the Carreau-Yasuda model~\cite{bird1987}.
The measured shear viscosity, $\eta$, was in the range 2~Pa-s~$\ge \eta \ge$~0.5~Pa-s for shear rates in the range $0.01 \le \dot{\gamma} \le 10$~s$^{-1}$.  \par
    
For flow experiments through pillar arrays, the viscoelastic fluid was pressure-driven through the microfluidic channels (Elveflow OB1), and video microscopy (Nikon Ti-e; 10$\times$, 0.3~NA objective) captured the motion (100 fps; Andor Zyla) of fluorescent tracer particles. 
Time-resolved velocity fields, $\textbf{u}(\textbf{r},t)$, were measured using PIV~\cite{Thielicke2014}, and Lagrangian statistics were obtained by simultaneous particle tracking.
A maximum Reynolds number of $\mathrm{Re} = \rho U d /\eta \lesssim 10^{-4}$ (density, $\rho$; mean flow speed, $U$) ensured that inertial effects were negligible and that the emergence of flow instability only depended on elastic effects. 
Experiments were limited to a transport regime dominated by advection, as determined by the P\'eclet number, $\mathrm{Pe} = U d / D(\eta)$, where $10^5 \le \mathrm{Pe} \le 10^9$ and $D$ accounts for the viscosity dependent Stokes-Einstein diffusion coefficient of the tracers. \par

\subsection{Numerical methods}
The numerical simulations were performed in a two-dimensional domain that was designed to exactly match the region-of-interest in experiments [Fig~\ref{fig:VEDisp_fig1}~(a)] via computer-generated photomasks. The flow of incompressible polymeric fluid in the interstitial region of the porous geometry is described by the conservation of mass and momentum as:
\begin{equation}\label{eqn:VEDIsp_1}
\nabla \cdot \mathbf{u}=0,
\end{equation}
\begin{equation}\label{eqn:VEDIsp_2}
\rho \left(\frac{\partial \mathbf{u}}{\partial t}+\mathbf{u}\cdot \nabla \mathbf{u} \right )=-\nabla p+\nabla \cdot \boldsymbol \sigma,
\end{equation}
where $\mathbf{u}$ and $p$ are the fluid velocity and pressure field, respectively. The total stress tensor $\boldsymbol{\sigma}$ is written as $\boldsymbol{\sigma}=\boldsymbol{\sigma_s}+\boldsymbol{\sigma_p}$, where $\boldsymbol{\sigma_s}$ and $\boldsymbol{\sigma_p}$ are the solvent and polymeric stress tensor, respectively. For the Newtonian solvent, $\boldsymbol{\sigma_s}$ is given as $ \boldsymbol{\sigma_s}=\eta_s(\nabla \textbf{u}+\nabla \textbf{u}^T)$, where $\eta_s$ is the solvent viscosity. We chose the FENE-P constitutive equation to calculate the polymeric stresses because it captures both elasticity and shear thinning behaviours as well as finite extensibility of polymeric chains \cite{Bird1987vol2, Bird1980}:     
\begin{equation}\label{eqn:VEDIsp_3}
\boldsymbol{\sigma}_p+\frac{\tau}{f}\overset{\nabla}{\boldsymbol{\sigma}}_p=\frac{b\eta_p}{f}(\nabla \mathbf{u}+\nabla \mathbf{u}^T)-\frac{D}{Dt}\left(\frac{1}{f}\right)[\tau \boldsymbol{\sigma}_p+b\eta_p \mathbf{I}],
\end{equation}
where $\tau$ is the polymeric chain relaxation time, and $\eta_p$ is the polymeric contribution to the zero shear viscosity of the solution, $\eta =\eta_s+\eta_p$. $\mathbf{I}$ is the identity tensor and $D/Dt$ is the material derivative. The function $f$ is described as:
\begin{equation}\label{eqn:VEDIsp_4}
f(\boldsymbol{\sigma_p})=\frac{L^2+\frac{\tau}{b\eta_p} tr(\boldsymbol{\sigma_p})}{L^2-3},
\end{equation}
where $b=L^2/(L^2-3)$, and the parameter $L^2=3R_0^2/R_e^2$ represents the ratio of the maximum allowable length, $R_0$, to the equilibrium length, $R_e$, of the polymeric chains \cite{ Bird1987vol2,Bird1980,Purnode1998}. 
For the  FENE-P model, a typical range of $L^2$ is 10-1000 \cite{Bird1980,Chilcott1988,Oliveira2002,Aramideh2019}, which reduces to the Oldroyd-B constitutive model in the limit of $L^2 \to \infty$. 
The upper convective time derivative operator $\nabla$ used in equation \ref{eqn:VEDIsp_3} is given by:
\begin{equation}\label{eqn:VEDIsp_5}
\overset{\nabla}{\boldsymbol{\sigma}}_p=\frac{D\boldsymbol{\sigma}_p}{Dt}-\boldsymbol{\sigma}_p\cdot\nabla\mathbf{u}-\nabla\mathbf{u}^T \cdot\boldsymbol{\sigma}_p.
\end{equation}

The numerical simulations were performed using an open-source framework RheoTool \cite{Pimenta2017} integrated with OpenFOAM \cite{Jasak2007}, where the equations were discretized using a finite volume method and the log-conformation approach was used to calculate the polymeric stress tensor. 
The relationship between the polymeric stress tensor, $\boldsymbol{\sigma_p}$, and the log-conformation tensor, $\mathbf{\Theta}$, is given as:
\begin{equation}\label{eqn:VEDIsp_6}
\boldsymbol{\sigma}_p=\frac{\eta_p}{\tau}(f e^{\mathbf{\Theta}}-b\mathbf{I}).
\end{equation}
The implementation and the validation of the numerical tool can be found in previous works~\cite{Pimenta2017,Favero2010}. 
The dimensionless numbers used in the simulations were: $\textrm{Re} = 10^{-4}$, $0.1 \le \textrm{Wi} \le 5$, $\beta=\eta_s/(\eta_s+\eta_p)=0.02$, and $L^2=1000$. The flow was driven with a constant inlet velocity of 50 $\mu$m/s on the left side of the channel with no-slip boundaries on the top and bottom walls [Fig~\ref{fig:VEDisp_fig1}]. \par

For both experiments and simulations, the Lagrangian stretching, $S(\mathbf{x},t)$, is determined from the time resolved flow field~\cite{Haller2015, voth2002} by using established methods~\cite{Parsa2011, Dehkharghani2019, Kumar2023PNAS}.
Briefly, fluid element positions, $\mathbf{x}_0$, at time $t=t_0$ are deformed by a flow field, $\mathbf{u}(\mathbf{x},t)$, and advected to new positions, $\mathbf{x}$, at time, $t_1=t_0+\lambda$.
The flow map, $\mathbf{F}_\lambda(\mathbf{x}_0) = \mathbf{x}$, is determined as the solution to $\frac{d\mathbf{x}}{dt} =\mathbf{u}(\mathbf{x}, t)$, and the (right) Cauchy-Green strain tensor is formed as $\mathbf{C}_\lambda^{(R)} = \nabla \mathbf{F}_\lambda^T \nabla \mathbf{F}_\lambda$.
Finally, the stretching field $S(\mathbf{x},t)$ is calculated as the square root of the largest eigenvalue of $\mathbf{C}_{\lambda}^{(R)}$, where the corresponding eigenvector gives the principal stretching direction. 
For all flows, the stretching history is determined by backward time integration, and the integration time was chosen as the fluid relaxation time ($\lambda = -\tau$)~\cite{Kumar2023PNAS}.
$\mathbf{F}_\lambda$ was determined through numerical integration (ODE45, MATLAB) for initial positions on a regular grid ($251 \times 351$) along with four auxiliary points each (1~$\mu$m separation). 
$\nabla \mathbf{F}_\lambda$ was computed through central differences of the auxiliary points.

\section{Results}
\subsection{Anisotropic Lagrangian transport in porous media}
\begin{figure*}
    \centering
    \includegraphics{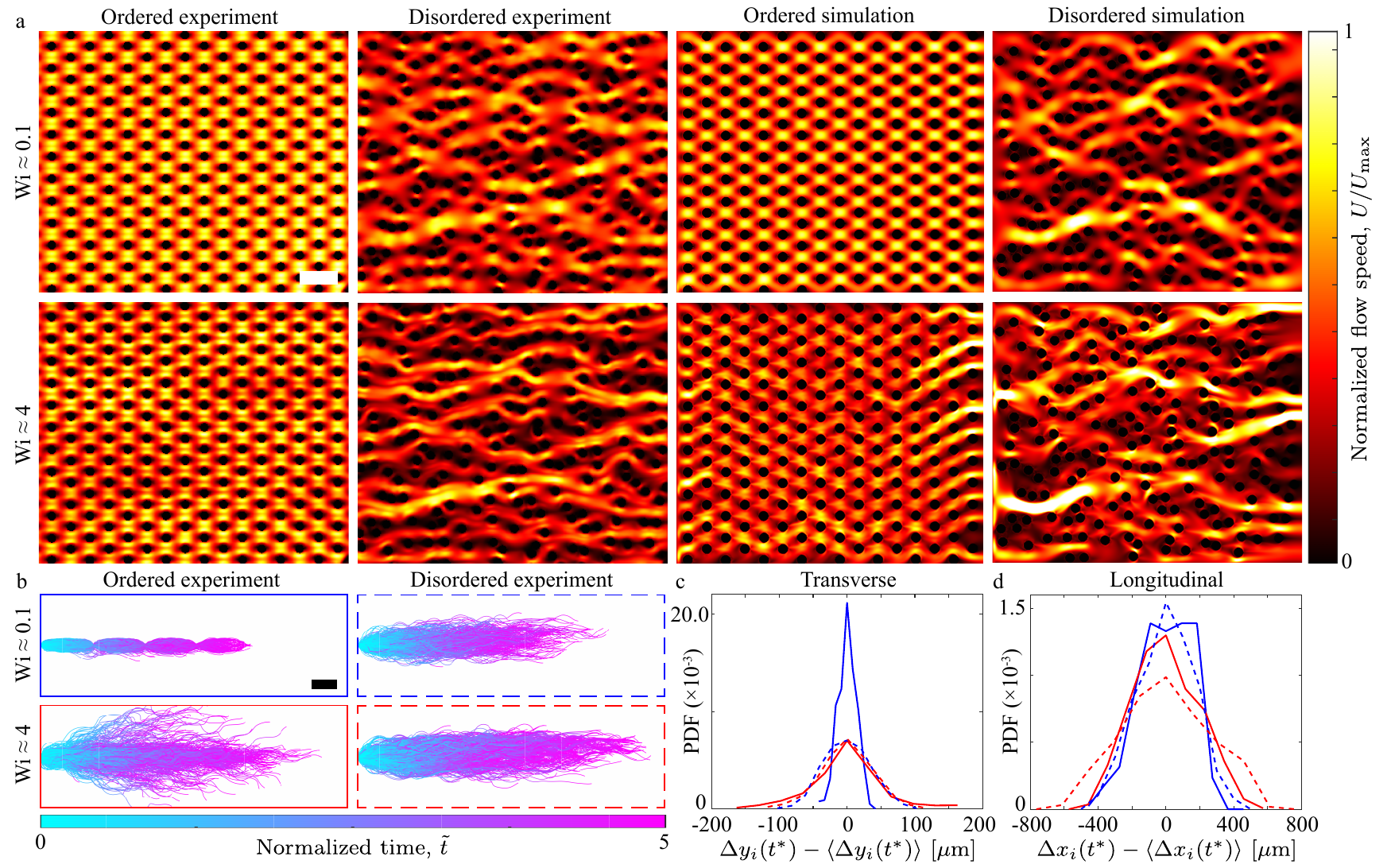}
    \caption{Experiments and simulations reveal Eulerian and Lagrangian transport of viscoelastic fluids in porous media. (a) Time-averaged flow fields for experiments and respective simulations at various Wi for ordered and disordered porous geometries. Flow is left to right. Scale bar, 120~$\mu$m. 
    (b) Experimentally measured particle trajectories with initial position subtracted, $\Delta\mathbf{x}_i(\tilde{t}) = \mathbf{x}_i(\tilde{t})-\mathbf{x}_i(0)$, for dimensionless time, $\tilde{t} = t U / l$, in the range $0 \le \tilde{t} \le 5$. Scale bar, $60~\mu$m. 
    (c)-(d) Transverse and longitudinal relative displacement distributions, respectively, $\Delta\mathbf{x}_i(t^\ast)-\langle\Delta\mathbf{x}_i(t^\ast)\rangle$, corresponding to (b) at normalized time, $t^\ast \equiv \tilde{t}=5$. Color and line type correspond to panel borders in (b).}
    \label{fig:VEDisp_fig1}
\end{figure*}

Flow through hexagonally ordered (staggered) geometries in both experiments and simulations exhibits a transition to unstable flow at a critical Weissenberg number~\cite{Walkama2020}, $\textrm{Wi}_\textrm{cr} \approx 0.5$. However, the time-averaged flow fields do not show a strong topological change with Wi due to the high degree of geometric symmetry [Fig.~\ref{fig:VEDisp_fig1}~(a)]. In contrast, disordered geometries stabilize these flows via the formation of preferential flow paths~\cite{Stoop2019}, where extensional fluid deformations -- and consequently polymer stretching -- are minimized~\cite{Walkama2020, Haward2021}. Time-averaged flow fields through disordered media in both experiments and simulations display a topological shift from a Newtonian flow [Fig.~\ref{fig:VEDisp_fig1}~(a)], where filaments form as Wi is increased and the flow field becomes more heterogeneous. As in previous works~\cite{Walkama2020,Haward2021}, this Eulerian picture points to a trade-off between stability and channelization that is mediated by pore microstructure. However, this framework provides little insight into the effect of geometry on fluid transport.\par

Turning toward a Lagrangian description of fluid transport reveals that lateral and longitudinal tracer displacements are enhanced at high Wi for the ordered and disordered porous geometries, respectively. 
Particle tracking provides tracer particle trajectories in time, $\mathbf{x}_i(\tilde{t})$, where $i$ represents an individual particle track. 
The normalized time, $\tilde{t}=tU/l$, corresponds to the number of pores traveled for a given characteristic (mean) flow speed, $U$, and stream-wise pore spacing, $l = a \sin 60^\circ$.
Examination of the net displacement of the fluid tracers with respect to their initial positions, $\Delta\mathbf{x}_i(\tilde{t}) = \mathbf{x}_i(\tilde{t})-\mathbf{x}_i(0)$, demonstrates how geometry influences transport through viscoelasticity [Fig.~\ref{fig:VEDisp_fig1}~(b)]. 
Tracers in the ordered geometry at small Wi tamely oscillate back and forth, as they weave through the pillar array following streamlines in the steady flow.
Conversely, at $\textrm{Wi} = 4 > \textrm{Wi}_\textrm{cr}$, tracers exhibit wild lateral excursions accompanied by a moderate enhancement of longitudinal displacement.
The former is consistent with previous observations of ``lane-changing''~\cite{Scholz2014}, which is a consequence of temporal velocity field fluctuations~\cite{Walkama2020}.
In the disordered geometry, tracers laterally explore a relatively large swatch of the porous channel by virtue of the meandering streamlines at low Wi, with little change at higher Wi [Fig.~\ref{fig:VEDisp_fig1}~(b)]. 
While the disordered flow remains steady at $\textrm{Wi} = 4$, tracer displacements are appreciably enhanced in the longitudinal direction.

The displacement distributions relative to the mean, $\Delta\mathbf{x}_i(t^*)-\langle\Delta\mathbf{x}_i(t^*)\rangle$, at a fixed time ($t^*\equiv\tilde{t}=5$) more clearly show the anisotropic enhancement of tracer excursions in both the ordered and disordered systems at high Wi [Fig.~\ref{fig:VEDisp_fig1}~(c)-(d)]. 
Here, $\langle \cdot \rangle$ indicates an ensemble average.
The transverse displacement [Fig.~\ref{fig:VEDisp_fig1}~(c)] exhibits a narrow distribution for low Wi in the ordered media (blue solid) due to the high P\'eclet number, stable flow.
At high Wi, the ordered media shows large displacements in the transverse direction (red solid), consistent with the onset of the elastic instability. 
Transverse disordered flow, on the other hand, is generally unaffected by increasing Wi and shows little to no change (blue and red dashed for low and high Wi, respectively). Surprisingly, longitudinal displacements [Fig.~\ref{fig:VEDisp_fig1}~(d)] show the opposite effect as a function of disorder. 
Longitudinal displacements in the ordered system (solid curves) show little change with Wi. 
However, the disordered media exhibits a broader tracer displacement distribution for high Wi (dashed red) than low Wi (dashed blue). 
Thus, tracers disperse by traveling both significantly faster and slower than the mean flow speed in disordered media at high Wi.
\par

\subsection{Mean squared displacement analysis reveals diffusive spreading of fluid tracers}
\begin{figure*}
    \centering
    \includegraphics{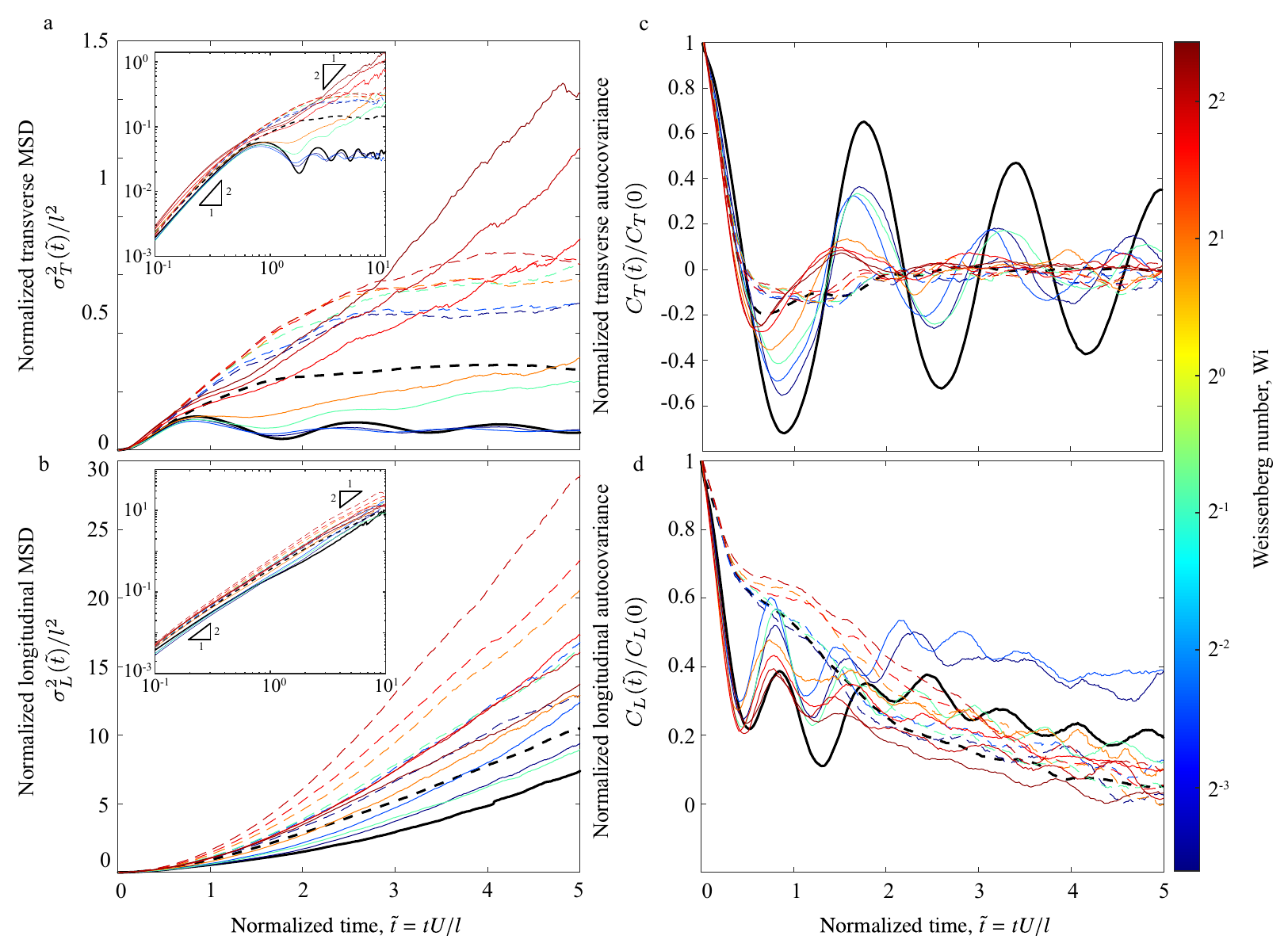}
    \caption{Experimental Lagrangian flow statistics inform the effect of geometry on tracer dispersal. (a)-(b) Normalized transverse and longitudinal mean squared displacement, respectively, as a function of normalized time for ordered (solid) and disordered (dashed) media at various Wi. Insets show the same data on a log-log scale. (c)-(d) Normalized transverse and longitudinal autocovariance, respectively, as a function of normalized time for ordered (solid) and disordered (dashed) flow at various Wi numbers. Black curves correspond to Newtonian fluid flow (100\% glycerol) at comparable low Re and high Pe numbers for all plots.
    }
    \label{fig:VEDisp_fig2}
\end{figure*}
The variance of the displacement distributions [Fig.~\ref{fig:VEDisp_fig1}~(c)-(d)] defines the advection-free mean squared displacement (MSD) at time, $\tilde{t}$, which indicate the nature and rate of spreading of the tracers:
\begin{align}
    \sigma^2_{T}(\tilde{t}) &= \left\langle \left(\Delta y_i(\tilde{t}) - \langle \Delta y_i(\tilde{t}) \rangle \right)^2 \right\rangle,\\
    \sigma^2_{L}(\tilde{t}) &= \left\langle \left( \Delta x_i(\tilde{t}) - \langle \Delta x_i(\tilde{t}) \rangle \right)^2 \right\rangle.
\end{align}
The MSD describes the average separation of fluid parcels from one another in time due to both the mechanical dispersion and flow instability in the transverse ($\sigma^2_{T}$) and longitudinal ($\sigma^2_{L}$) directions, respectively. 
The transverse MSD in the ordered media exhibits oscillations at low Wi reflective of the obstacle periodicity, but the MSD saturates due to the sampling of streamlines with finite amplitude displacements from the mean flow direction.
As Wi increases and the flow becomes elastically unstable [Fig.~\ref{fig:VEDisp_fig2}~(a), solid curves], the displacements at long times are unbounded and grow superlinearly in time [Fig.~\ref{fig:VEDisp_fig2}~(a), inset], which is indicative of superdiffusive transport (i.e., $\textrm{MSD}\propto t^\alpha$ with $1<\alpha<2$).
However, in the disordered geometry, all transverse MSDs plateau after $\approx2-3$ pore lengths and are only mildly affected by Wi [Fig.~\ref{fig:VEDisp_fig2}~(a), dashed curves]. While all transverse MSDs are ballistic at short times ($\alpha \approx 2$) [Fig.~\ref{fig:VEDisp_fig2}~(a), inset], only MSDs for the transverse, high Wi ordered flow continue growing superlinearly at long times due to the elastic instability.
Mechanical dispersion in the flow direction~\cite{Taylor1953, Aris1956} causes longitudinal MSDs to be unbounded. While all longitudinal MSDs are superdiffusive  [Fig.~\ref{fig:VEDisp_fig2}~(b), inset], MSDs for viscoelastic flows grow faster in the disordered media [Fig.~\ref{fig:VEDisp_fig2}~(b), dashed curves] compared to ordered media [Fig.~\ref{fig:VEDisp_fig2}~(b), solid curves], relative to their respective Newtonian flows.

\subsection{Dispersion tensor for viscoelastic porous media flow}

The average rate at which a tracer ensemble spreads is parameterized by the dispersion coefficient. In the present case, due to the  observed anisotropic transport, we examine the time-dependent dispersion tensor~\cite{maier2000}, which we compare across experiments and simulations:
\begin{equation}
    D_{T, L}(\tilde{t}) = \int_0^{\tilde{t}} C_{T,L}(t^\prime)dt^\prime.
    \label{eqn:dispersioncoeff}
\end{equation}
Here, $C_{T,L}(\tilde{t})$ is the time-dependent velocity autocovariance that quantifies the temporal correlation of tracer velocity:
\begin{align}
    C_{T}(\tilde{t}) = \frac{1}{N}\sum_{i=1}^N \left(u_{yi}(\tilde{t})-\langle u_{yi}(\tilde{t})\rangle\right)\left(u_{yi}(0)-\langle u_{yi}(0)\rangle\right),\\
    C_{L}(\tilde{t}) = \frac{1}{N}\sum_{i=1}^N \left(u_{xi}(\tilde{t})-\langle u_{xi}(\tilde{t})\rangle\right)\left(u_{xi}(0)-\langle u_{xi}(0)\rangle\right),
\end{align}
where the $u_{yi}(\tilde{t})$ and $u_{xi}(\tilde{t})$ are respectively the transverse and longitudinal velocity components of particle $i$, and $\langle \cdot \rangle$ is an ensemble average over $N$ particles.
The transverse autocovariance is periodic about zero in the ordered geometry, but the oscillations lose coherence with the onset of the instability as Wi increases.
However, as the instability drives lane changing, transverse particle velocities gain a slight net correlation due to the motion over one or more pores lateral to the flow. 
For disordered flow, $C_T$ rapidly decays and appears to have little dependence on Wi [Fig.~\ref{fig:VEDisp_fig2}~(c)], due to the random uncorrelated flow paths through the medium. 
Similar to the transverse direction, the longitudinal autocovariance for ordered flow exhibits periodic peaks that lose coherence as Wi increases [Fig.~\ref{fig:VEDisp_fig2}~(d)], due to the onset of spatiotemporal velocity fluctuations. 
Conversely, $C_L$ for disordered flow shows an increase in the correlation time beyond a Newtonian fluid as Wi is increased. 
This increased velocity correlation stems from the formation of preferential flow paths that transport fluid in the longitudinal direction~\cite{Walkama2020}, which we expect to lead to increased dispersion. \par
\begin{figure*}
    \centering
    \includegraphics{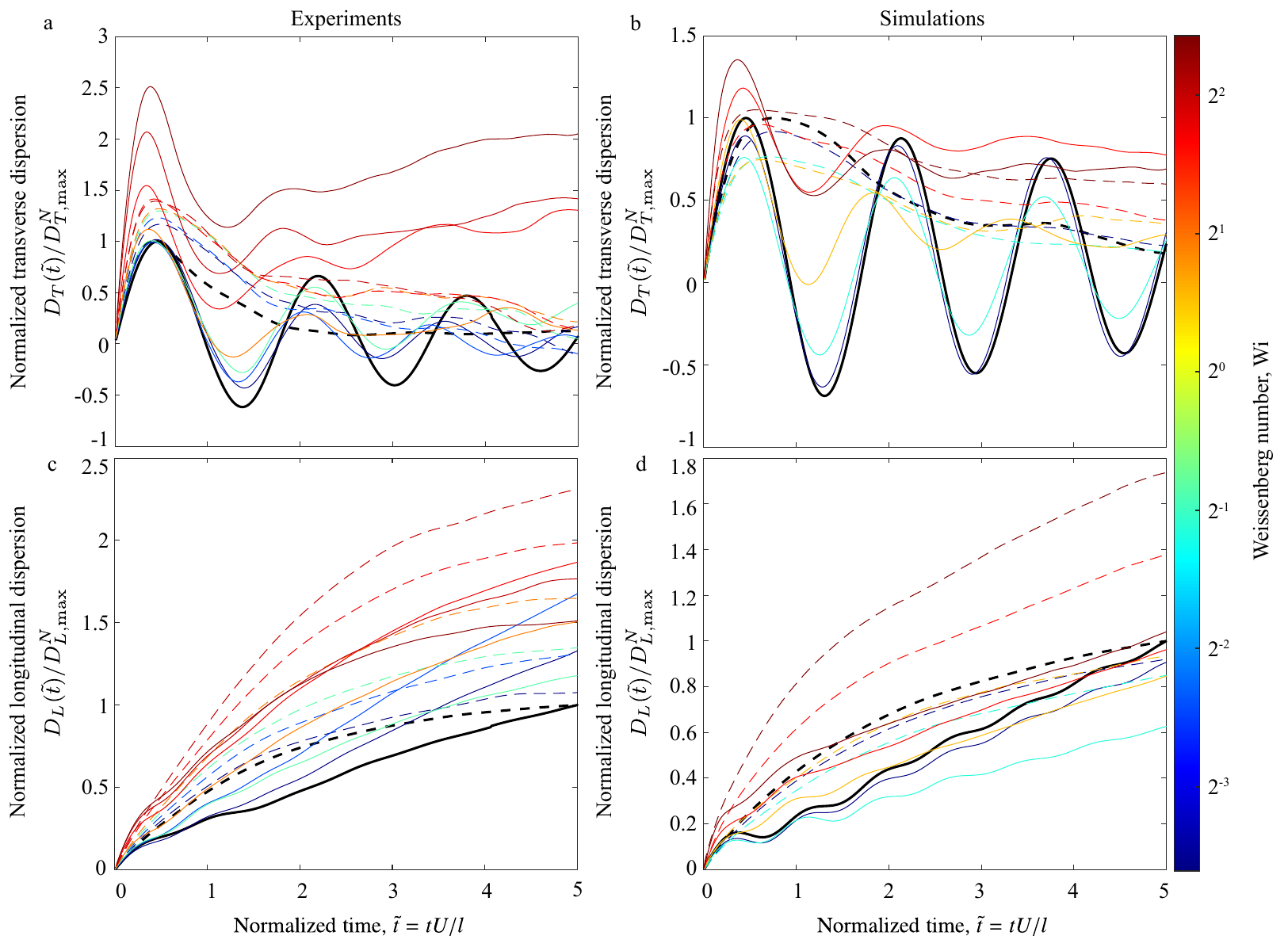}
    \caption{Comparison of dispersion tensor components for experiments and simulations in ordered (solid) and disordered (dashed) media. 
    (a)-(b) Transverse dispersion coefficients for experiments and simulations, respectively. 
    (c)-(d) Longitudinal dispersion coefficients for experiments and simulations, respectively. 
    Values for all dispersion coefficients are normalized by their respective maximal Newtonian values for ordered and disordered media in the longitudinal and transverse direction across experiments and simulations.}
    \label{fig:VEDisp_fig3}
\end{figure*}
The time-dependent dispersion tensor serves as a primary measure of augmented anisotropic transport.
In these high P\'eclet flows, transverse dispersion coefficients in both experiments [Fig.~\ref{fig:VEDisp_fig3}~(a)] and simulations [Fig.~\ref{fig:VEDisp_fig3}~(b)] either oscillate about zero (ordered) or decay to zero (disordered) for stable flows ($\textrm{Wi} \lesssim \textrm{Wi}_{cr}$). 
In experiments, the instability only occurs in the ordered flow [Fig.~\ref{fig:VEDisp_fig3}~(a), solid curves].
However, in simulations, some velocity fluctuations in both ordered and disordered flows cause finite values of $D_T$ at long times and high Wi [Fig.~\ref{fig:VEDisp_fig3}~(b)]. 
While both geometries ultimately become unstable in simulations, this effect is more pronounced in high Wi, ordered flows, indicated by elevated long-time dispersion [Fig.~\ref{fig:VEDisp_fig3}~(b)] compared to the disordered flow. 
In the longitudinal direction, $D_L$ grows approximately linearly for ordered flow at low Wi, but plateaus at high Wi [Fig.~\ref{fig:VEDisp_fig3}~(c), solid curves], indicating an effectively diffusive regime. 
The linear growth is due to the constant, non-zero autocovariance in steady flows through the ordered geometry. 
Once the flow becomes unstable in experiments ($\textrm{Wi} \gtrsim \textrm{Wi}_{cr}$), dispersion values increase at small times but plateau at long times [Fig.~\ref{fig:VEDisp_fig3}~(c), solid curves] due to decorrelation of observed in $C_L$ [Fig.~\ref{fig:VEDisp_fig2}~(d), solid curves]. 
This effect is also seen in simulations [Fig.~\ref{fig:VEDisp_fig3}~(d), solid curves], but the dispersion coefficient does not reach a constant value in time. 
In simulations, we observe far less diagonal flow due to the reduced system size compared to experiments. 
As a result of this correlated longitudinal motion, $D_L$ does not reach a steady state in simulations. 
Due to measurement limitations in experiments, it is unclear when high Wi disordered flow reaches a steady state [Fig.~\ref{fig:VEDisp_fig3}~(c), dashed curves]. However, the magnitude of dispersion is larger than ordered flows for the same Wi at all times. 
This observation also holds true for simulations [Fig.~\ref{fig:VEDisp_fig3}~(d), dashed curves], where $ D_L $ disordered media is much larger than for ordered media and is still growing for long times.
\par

We have systematically quantified the transport properties and stability of viscoelastic flows through porous media.
These results show that there is a clear trade-off between rheology-enhanced transport and geometry that also accompanies the stabilizing effect of preferential flow paths in the disordered media.
These viscoelastic fluids are composed of elastic polymers suspended in a carrier fluid, which stretch dramatically in extensional strain compared to shear strain~\cite{Smith1999}. Polymer elasticity embeds a memory into the fluid, whereby polymers continue to stretch and accumulate stress as they move through both space and time via advection. Flow through the staggered ordered media becomes chaotic due to extensional stresses at high Wi, whereas preferential flow paths can alleviate extension and promote stability in the disordered media~\cite{Walkama2020,Haward2021}. The topology of the polymeric stress field has been known to regulate flow states in viscoelastic flows \cite{Kumar2021multistability,Kumar2021tristability, Kumar2022hysteresis, Mokhtari2022}. Polymeric stress is thus integral to both enhanced transport and elastic stability in these systems, but access to the polymeric stress field in experiments is challenging \cite{Sun2016}. Recently, a Lagrangian analysis of fluid deformation was demonstrated to provide direct insight into the topology of the polymeric stress field from readily measurable flow field data \cite{Kumar2023PNAS}, which is a key to the comprehensive understanding of the transport in these systems.


\subsection{Lagrangian stretching guides fluid flow}
Lagrangian coherent structures (LCS)~\cite{Haller2015} characterize material lines that organize fluid transport, which have been applied broadly across scales to understand ocean flow patterns, chaotic mixing~\cite{voth2002}, bacterial transport flows~\cite{Dehkharghani2019}, and complex fluid flows~\cite{Kumar2023PNAS}. 
Key to LCS analysis is the concept of the Lagrangian fluid stretching field, which quantifies the extensional strain history of fluid elements and is closely linked to the finite-time Lyapunov exponent (FTLE) field. 
Manifolds of the stretching field act as barriers to advective transport, and recently, were shown to be highly correlated with the topology of the polymeric stress field in viscoelastic flows~\cite{Kumar2023PNAS}.
Thus, the Lagrangian stretching could potentially provide a direct link between the polymer stress and dispersion for the viscoelastic porous media flows considered here. \par

\begin{figure*}
    \centering
    \includegraphics{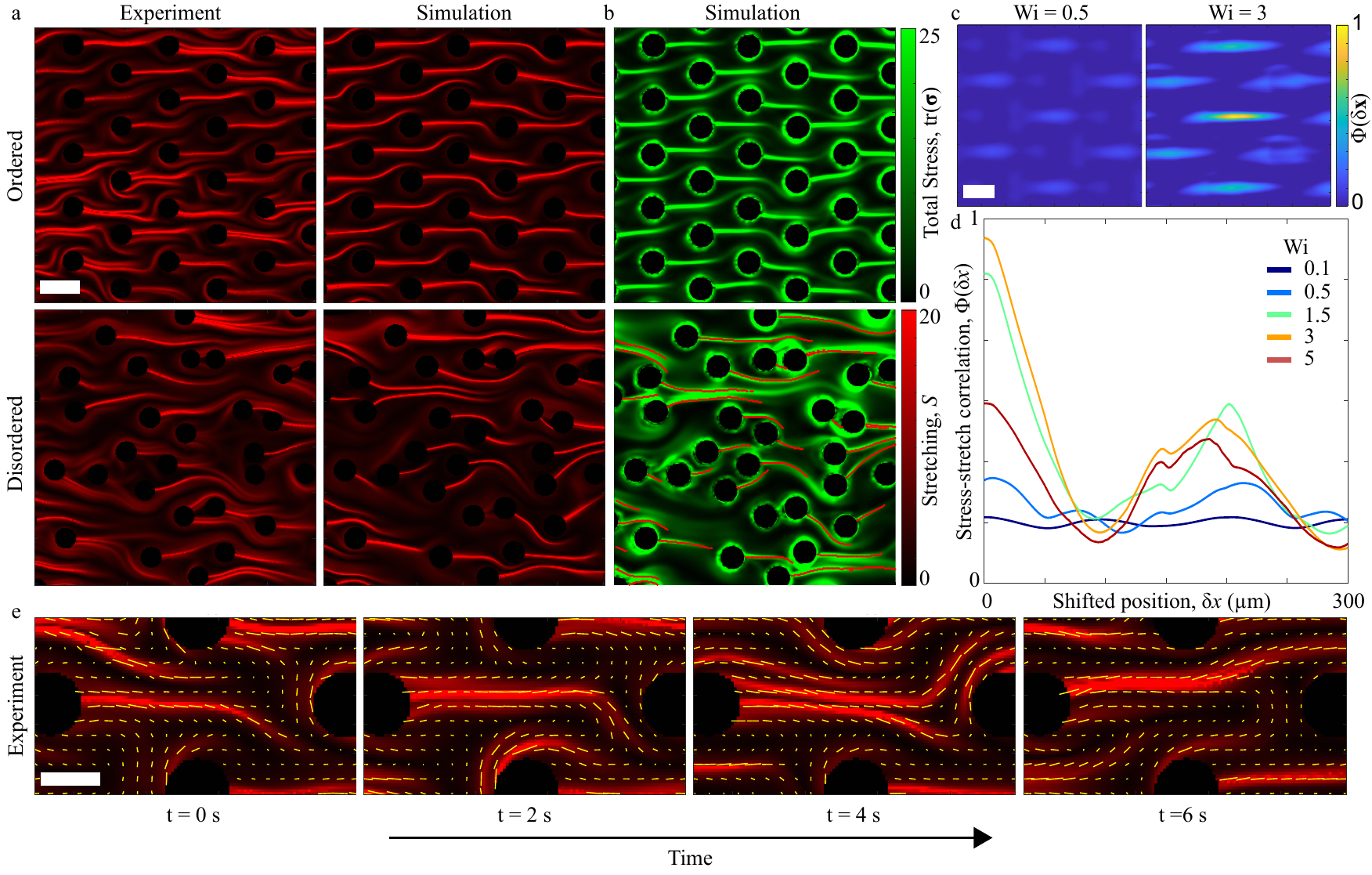}
    \caption{Lagrangian stretching correlates to stress and illustrates the pathways guiding fluid transport. 
    (a) Stretching fields for both the ordered and disordered geometries in experiments and simulations at Wi = 3. Scale bar, 100 $\mu$m. 
    (b) Stress fields from simulations for corresponding flows in panel (a). Magenta lines show manifolds of the stretching field, which correspond to ridges of maximum stretching. 
    (c) Stress-stretch correlation maps for simulated, ordered flow fields at Wi = 0.5 (left) and Wi = 3 (right). Center corresponds to $\delta\mathbf{x}=0$. Scale bar, 50 $\mu$m. 
    (d) Stress-stretch correlation for different shift positions in the longitudinal ($x$) direction  for simulated flow fields at various Wi ($\delta y = 0$). 
    (e) Time series of experimental stretching fields showing a wavering stretching manifold in ordered porous media at Wi = 3. Vector fields indicate the principle stretching direction. Scale bar, 50 $\mu$m.
    }
    \label{fig:VEDisp_fig4}
\end{figure*}

Lagrangian stretching fields, $S(\textbf{x}, t)$, were calculated directly from both the experimentally measured and simulated velocity fields (see Methods)~\cite{Parsa2011,Haller2015, Dehkharghani2019,Kumar2023PNAS}.
The viscoelastic fluid relaxation time, $\tau$, was chosen as a natural integration time over which the stretching history was computed for all Wi. 
Stretching fields for both the ordered and disordered media [Fig.~\ref{fig:VEDisp_fig4}~(a)] reveal sharp regions of high stretching (manifolds) that generally emanate from the hyperbolic flow regions on the downstream sides of the pillars including unsteady flow conditions [Fig.~\ref{fig:VEDisp_fig4}~(e)].
Importantly, simulations also provide the time-dependent stress tensor, ${\boldsymbol \sigma}(\mathbf{x},t)$, and enable direct comparison with the Lagrangian stretching [Fig.~\ref{fig:VEDisp_fig4}~(b)]. 
In line with recent work \cite{Kumar2023PNAS}, the trace of the polymeric stress tensor [Fig.~\ref{fig:VEDisp_fig4}~(b)] mirrors the topology of the stretching manifolds [Fig.~\ref{fig:VEDisp_fig4}~(a)] for both ordered and disordered flow simulations.
To quantify the correlation between the topologies of stress and stretching fields, the cross-correlation is defined as:
\begin{align*}
    \Phi(\delta \mathbf{x}) &= \frac{[ \textrm{tr}({\boldsymbol \sigma}(\mathbf{x}+\delta \mathbf{x})) - \langle \textrm{tr}({\boldsymbol \sigma}) \rangle ] \cdot [S(\mathbf{x})- \langle S \rangle]}
    {\langle \textrm{tr}({\boldsymbol \sigma}) \rangle \langle S \rangle},
\end{align*}
where $\langle \cdot \rangle$ denotes the mean value over all $\mathbf{x}$, and $\delta \mathbf{x}$ is the shifted position. 
Large values of $\Phi$ indicate large values for both $\textrm{tr}({\boldsymbol \sigma})$ and $S$. 
The strongest correlation occurs for $\delta\mathbf{x} = 0$ due to the overlap of the filamentous stretching and stress fields [Fig.~\ref{fig:VEDisp_fig4}~(c)], which is indicated by the elongated features of $\Phi$ for the ordered media in the longitudinal direction. 
Examining the magnitude of the cross-correlation in the flow direction ($\delta x$) shows that the strength of the of $\Phi$ initially increases with Wi before diminishing at larger Wi [Fig.~\ref{fig:VEDisp_fig4}~(d)], likely due to the onset of strong temporal fluctuations.
\par

The concordance between Lagrangian stretching and stress demonstrated by simulations provides insight into the role of stress in dispersive transport for viscoelastic flow experiments in porous media~\cite{Kumar2023PNAS}.
In experiments for ordered media, enhanced transverse dispersion is driven by elastic instability.
The accompanying mobility of stretching manifolds [Fig.~\ref{fig:VEDisp_fig4}~(e)] -- which act as barriers to advective transport -- effectively guide the local flow.
A time series of stretching fields from ordered experiments at high Wi shows that stretching manifolds span the pillar array in the longitudinal direction.
Their lateral fluctuations in the transverse direction illustrate the mechanism of lane-changing~\cite{Scholz2014} in enhanced transverse dispersion. 
Conversely, stable flows in disordered media at high Wi disallow transverse material flux.
In this case, stretching manifolds elongate as Wi increases and cut off regions of locally high pillar density [Fig.~\ref{fig:VEDisp_fig4}~(a)]. 
The stretching manifolds thus enhance longitudinal transport by acting as a conduit through the porous media.
In sum, instability in ordered media allows these stretching manifolds to mobilize, whereas the stability in disordered media forces the stretching manifolds, and therefore the stress, to cut off regions of flow that were previously available at low Wi.

\section{Conclusions}
In this work, we quantify dispersion in viscoelastic flow through porous media and show how it is driven by viscoelastic instabilities. 
Flow through hexagonally-ordered (staggered) porous media enhances transverse dispersion, which is especially prevalent at high Weissenberg number.
Through a novel Lagrangian analysis of fluid stretching fields, we illustrate that this phenomenon is regulated by stretching manifolds that act as barriers to advective transport and characterize high-stress regions of the flow. 
At low Wi, material lines are symmetric, stable, and situated between obstacles parallel to flow, while at high Wi strong lateral fluctuations guide lateral dispersion. 
Conversely, high Wi flows in disordered porous media remain stable, as previously shown, due to the reduction of extension and the availability of preferential flow paths~\cite{Walkama2020, Haward2021}. 
This flow stability is reflected in stable stretching manifolds that disallow random transverse flows. 
Stretching manifolds also increase spatial flow speed heterogeneity by cordoning off slow flow regions as the stretching manifolds elongate with increasing Wi. 
This feature results in dead zones between high-speed filaments, where flow is carried longitudinally at disparate respective rates. 
Thus, these results show that a Lagrangian examination of fluid stretching is essential to gain insight into the coupling between fluid transport and stress through mechanical fluid instability in viscoelastic porous media flows. 

\section*{Acknowledgments}
The authors acknowledge support for this work by National Science Foundation awards CBET-2141349, CBET- 1701392, and CAREER-1554095 (to J.S.G.), and CBET-1700961, CBET-1705371, and CBET-2141404 (to A.M.A.).


\providecommand{\noopsort}[1]{}\providecommand{\singleletter}[1]{#1}%

\end{document}